\documentclass[draftcls,onecolumn,12pt]{IEEEtran}
\usepackage{amssymb}
\usepackage{amsmath}
\usepackage{stfloats}
\usepackage{graphicx}
\usepackage{subfigure}
\usepackage{tabularx}
\usepackage{epsfig,epsf,color,balance,cite}
\usepackage{multirow}

\usepackage{dsfont}

\usepackage{amsthm}

\usepackage[T1]{fontenc}
\usepackage{aecompl}
\usepackage{color}

\newtheorem{Theorem}{Theorem}
\newtheorem{Lemma}{Lemma}

\begin{document}
% paper title
\title{Bandit Inspired Beam Searching Scheme for mmWave High-Speed Train Communications}

\author{
Jun-Bo Wang, Ming Cheng, Jin-Yuan Wang, Min Lin,\\Yongpeng Wu, Huiling Zhu, Jiangzhou Wang
\thanks{Jun-Bo Wang and Ming Cheng are with National Mobile Communications Research Laboratory, Southeast University, 210096 Nanjing, China. Email: \{jbwang, mingcheng\}@seu.edu.cn}
\thanks{Jin-Yuan Wang and Min Lin are with the Key Lab of Broadband Wireless Communication and Sensor Network Technology, Ministry of Education, Nanjing University of Posts and Telecommunications, Nanjing 210003, China. Email: \{jywang,linmin\}@njupt.edu.cn}
\thanks{Yongpeng Wu is with Shanghai Key Laboratory of Navigation and Location Based Services, Shanghai Jiao Tong University, Minhang 200240, China. Email: yongpeng.wu@sjtu.edu.cn}
\thanks{Huiling Zhu and Jiangzhou Wang are with the School of Engineering and Digital Arts, University of Kent, Canterbury, Kent, CT2 7NT, U.K. Email:  \{h.zhu, j.z.wang\}@kent.ac.uk}
%\thanks{Xiang-Gen Xia is with the Department of Electrical and Computer Engineering, University of Delaware, Newark, DE 19716 USA. E-mail: xxia@ee.udel.edu }
%\thanks{Kai-Kit Wong is with the Department of Electronic and Electrical Engineering, University College London, London WC1E 6BT, UK. Email: kai-kit.wong@ucl.ac.uk}

}

% make the title area
\maketitle
\begin{abstract}
High-speed trains (HSTs) are being widely deployed around the world. To meet the high-rate data transmission requirements on HSTs, millimeter wave (mmWave) HST communications have drawn increasingly attentions. To realize sufficient link margin, mmWave HST systems employ directional beamforming with large antenna arrays, which results in that the channel estimation is rather time-consuming. In HST scenarios, channel conditions vary quickly and channel estimations should be performed frequently. Since the period of each transmission time interval (TTI) is too short to allocate enough time for accurate channel estimation, the key challenge is how to design an efficient beam searching scheme to leave more time for data transmission. Motivated by the successful applications of machine learning, this paper tries to exploit the similarities between current and historical wireless propagation environments. Using the knowledge of reinforcement learning, the beam searching problem of mmWave HST communications is formulated as a multi-armed bandit (MAB) problem and a bandit inspired beam searching scheme is proposed to reduce the number of measurements as many as possible. Unlike the popular deep learning methods, the proposed scheme does not need to collect and store a massive amount of training data in advance, which can save a huge amount of resources such as storage space, computing time, and power energy. Moreover, the performance of the proposed scheme is analyzed in terms of regret. The regret analysis indicates that the proposed schemes can approach the theoretical limit very quickly, which is further verified by simulation results.
\end{abstract}


\IEEEpeerreviewmaketitle

\newpage

\section{{Introduction}}
\label{section1}
As a fast, convenient and green public transportation system, high-speed trains (HSTs) with a speed more than 300km/h are being deployed rapidly around the world. Characterized by high mobility, the HST scenario is expected to be a typical scenario in the fifth generation (5G) mobile communications and becomes an important topic \cite{HST01}. To enable intelligent transport system, numerous high-data-rate applications are required in this scenario, for instance, on-board and trackside high definition (HD) video surveillance, train operation information, real-time train dispatching HD video, and journey information \cite{HST02,HST03}. To guarantee the safety and quality-of-service (QoS) of these applications, a reliable communication link between the train and the ground plays a key role. The dominant wireless communication system for current railways is the global system for mobile for rail (GSM-R), which is a narrow-band communication system and only supports a peak data rate of 200kbps \cite{HST04}. Moreover, the GSM-R is specifically used for train control and cannot support high-data-rate transmissions. Using existing technologies, various schemes, such as the multimodal global mobile broadband communication (MOWGLY) developed by the Europe and the HST "Series N700" proposed in Japan, have been proposed for broadband wireless access on HSTs. However, the data transmission rates of current technologies are relatively low (e.g. 2~4Mbps), which is insufficient for the rapid growth of railway services. To meet the growing requirements of high-data-rate transmission, long-term evolution for railway (LTE-R) is recommended to provide broadband wireless communications for HSTs. However, due to the limitation of spectrum bandwidth, LTE-R cannot provide broadband radio access for each user in HSTs. Moreover, the spectrum used by current mobile communication system, which concentrates in sub-6 GHz, has become increasingly congested. To tackle this challenge, one approach is to adopt unlicensed spectrum bands. In recent years, how to exploit the mmWave band has drawn intensive attention in the academia and industry \cite{HST05,HST06}.

Compared to the sub-6 GHz signal, the millimeter wave (mmWave) signal has a wide bandwidth and a small wavelength. Since expanding the bandwidth is an effective and efficient approach to increase the system throughput, mmWave communication is a promising technology for the 5G \cite{HST32,HST35}. Moreover, the small wavelength of mmWave signals enables large antenna arrays to be placed in a compact size, which can provide high gains and directivities to overcome the inherent mmWave weakness, such as severe propagation loss and penetration loss. So far, many efforts have been devoted to the research and development of mmWave communications, and it has shown that mmWave communications can boost the transmission performance in indoor and outdoor scenarios. However, how to exploit the mmWave band in HST communications has been becoming a challege in the academia and industry \cite{HST07,HST08}.

\subsection{Previous works}
\label{section1a}

In conventional multiple-input multiple-output (MIMO) systems, each antenna is associated with an individual radio frequency (RF) chain such that fully digital beamforming can be implemented. However, in mmWave MIMO systems with large antenna arrays, due to the hardware limitations, such as hardware cost, complexity and power consumption, fully digital beamforming is infeasible \cite{HST09}. Moreover, thanks to the high penetration and propagation loss of mmWave signals, the mmWave channel is relatively sparse in the spatial domain. Thus, it is not necessary to implement fully digital beamforming while the number of propagation paths in the mmWave channel is limited \cite{HST10}. Therefore, analog beamforming \cite{HST11} and hybrid beamforming \cite{HST12} are usually adopted for mmWave communications, where phase shifters are often used to adjust the phase of the transmitted or received signal on each antenna to realize beamforming. Due to the additional hardware constraints and differences in beamforming architectures, classic channel estimation schemes used in traditional MIMO systems are not applicable for mmWave communication systems.

In mobile communications, channel estimation plays a key role to realize high quality services. Thanks to the sparsity of mmWave channels, the propagation paths of mmWave communications are normally estimated by searching the beam-steering directions of each path. Accordingly, beam training based algorithms were proposed in \cite{HST10,HST13}. Usually, these algorithms divide the process of finding each propagation path into multiple stages and refine the estimated angular range to narrow the beam patterns in each stage. However, these multi-stage algorithms can only estimate the channel state information (CSI) efficiently in point-to-point mmWave scenario. For multiple-user scenarios, the training overhead scales linearly with the number of users, which makes these multi-stage algorithms infeasible. To speed up the channel estimation process in multi-user scenarios, several multi-user beamforming based algorithms have been proposed to perform the simultaneous channel estimation for multiple channels. In \cite{HST14}, the frequency tone-based estimation algorithm was proposed for multi-user channels. In \cite{HST15}, the random compressed sensing-based estimation algorithms were analyzed by using random beam directions. The training overhead is reduced to a certain degree by predetermining the number of channel measurements performed by these random beamforming based algorithms. However, different numbers of measurements are required for users with different coherent time and signal-to-noise ratios (SNRs). Therefore, these algorithms may not work for all users so that they cannot achieve the optimal performance. In \cite{HST16}, the authors proposed a simultaneous-estimation with iterative fountain training (SWIFT) framework for the channel estimation of mutli-user mmWave MIMO systems, and the training duration required for estimating the multi-user channels is adaptively increased until a predetermined stopping criterion has been met at each user. By examining all these channel estimation schemes, it can be found that only the instantaneous measurements of CSI are exploited. Obviously, it is time-consuming to carry out the channel estimation of mmWave communications with a large number of antennas. Especially, in the HST scenarios, the mmWave channel varies quickly and the period of each transmission time interval (TTI) is too short to allocate enough time for accurate channel estimation. Thus, the existing schemes are not applicable for HST scenarios due to long estimation period.

In practice, the communication system is deployed and operated in a specific location, which indicates that the system performance can be optimized dynamically according to the measurements of wireless propagation environments. Recent investigations have found that these measurements convey a lot of similarities between current and historical wireless propagation environments \cite{HST37}. With the fast development of machine learning, it is very promising to exploit such similarities to further improve the communication performance. Out-of-band side information, such as data from sensors and other communications and congruency between the channels at mmWave and sub-6 GHz bands, are leveraged to assist beam training and channel estimation in \cite{HST17} and \cite{HST18}, respectively. In \cite{HST19}, the authors proposed a framework of generating 5G MIMO dataset using ray tracing, and deep learning model was applied to assist the mmWave beam selection. In \cite{HST20}, a deep learning solution was proposed for highly-mobile mmWave applications with coordinated beamforming. In \cite{HST21}, a deep learning based channel estimation algorithm was proposed for beamspace mmWave massive MIMO systems. It can be observed that applying deep learning in mmWave communication systems has drawn increasingly attentions. However, note that a massive amount of data must be collected and stored in advance for deep learning methods. In most cases, the learning process consumes a huge amount of computing time and power energy. Moreover, the learning process must be carried out repeatedly with the change of the scenarios. Therefore, although deep learning methods are promising to improve the performance of communication systems significantly, it is difficult to realize deep learning methods in practical systems over diverse scenarios, which motivate us to further explore alternative learning techniques to address the channel estimation in mmWave HST communication systems.

\subsection{Motivations and contributions}
\label{section1b}

The idea that we learn by interacting with our environment is probably the first to occur to us when we think about the nature of learning, which motivates the reinforcement learning. Reinforcement learning is to learn what to do and how to map situations to actions so as to maximize a numerical reward signal. The learner is not told which actions to take, but instead must figure out which actions yield the most reward by trying them. In other words, reinforcement learning is a method of machine learning in which data becomes available in a sequential order and is used to update the best action for future environments at each step. Compared with that of deep learning methods, the idea of reinforcement learning is attractive for the future communication systems, which can save the storage space, computing resource and power consumption significantly while boosting the transmission performance greatly.

In the HST scenarios, channel conditions vary quickly and channel estimations must be performed frequently. Since the period of TTI is very short, the number of measurements is constrained for each TTI, and it is very important to select which paths to be measured in the beam searching process. In reinforcement learning, this problem can be classified into the sequential and selection problems, which can be solved by the multi-armed bandit (MAB) \cite{HST27}. In the classic MAB, there are independent arms and a single player. At each time, the player chooses one arm to play and obtains a random reward over time from an unknown distribution. Different arms may have different reward distributions. The design objective is a sequential arm selection policy that maximizes the total expected reward over a horizon of length. Inspired by the successful applications of MAB in \cite{HST28,HST29,HST30}, this paper will try to formulate the beam searching problem of mmWave HST communications as a MAB problem. All possible paths can be modeled as arms and a bandit inspired beam searching scheme is proposed. During the channel estimation, the achievable transmission rates over the selected arms (i.e., paths) are measured and feed-backed as rewards. By updating the rewards of all arms, the wireless propagation environments can be learned and exploited for improving the accuracy of channel estimations while shortening the period of channel estimations. The performance of the proposed scheme is analyzed in terms of regret. The regret analysis indicates that the proposed schemes can approach the theoretical limit very quickly, which is further verified by simulation results.

The rest of this paper is organized as follows. The system model is introduced in Section~\ref{section2}. In Section~\ref{section3}, the beam searching problem is formulated for mmWave HST systems. The bandit inspired beam searching scheme is proposed in Section~\ref{section4} and the regret of the proposed scheme is analyzed in Section~\ref{section5}. Numerical results are presented in Section~\ref{section6}, and conclusions are drawn in Section~\ref{section7}.

\section{System model}
\label{section2}

In this section, we present our system model together with necessary assumptions, such as the network configuration, mmWave MIMO HST channel model and signal model.

\subsection{Network configuration}
\label{section2a}
\begin{figure}[ht]
\centering
\includegraphics[width=6in]{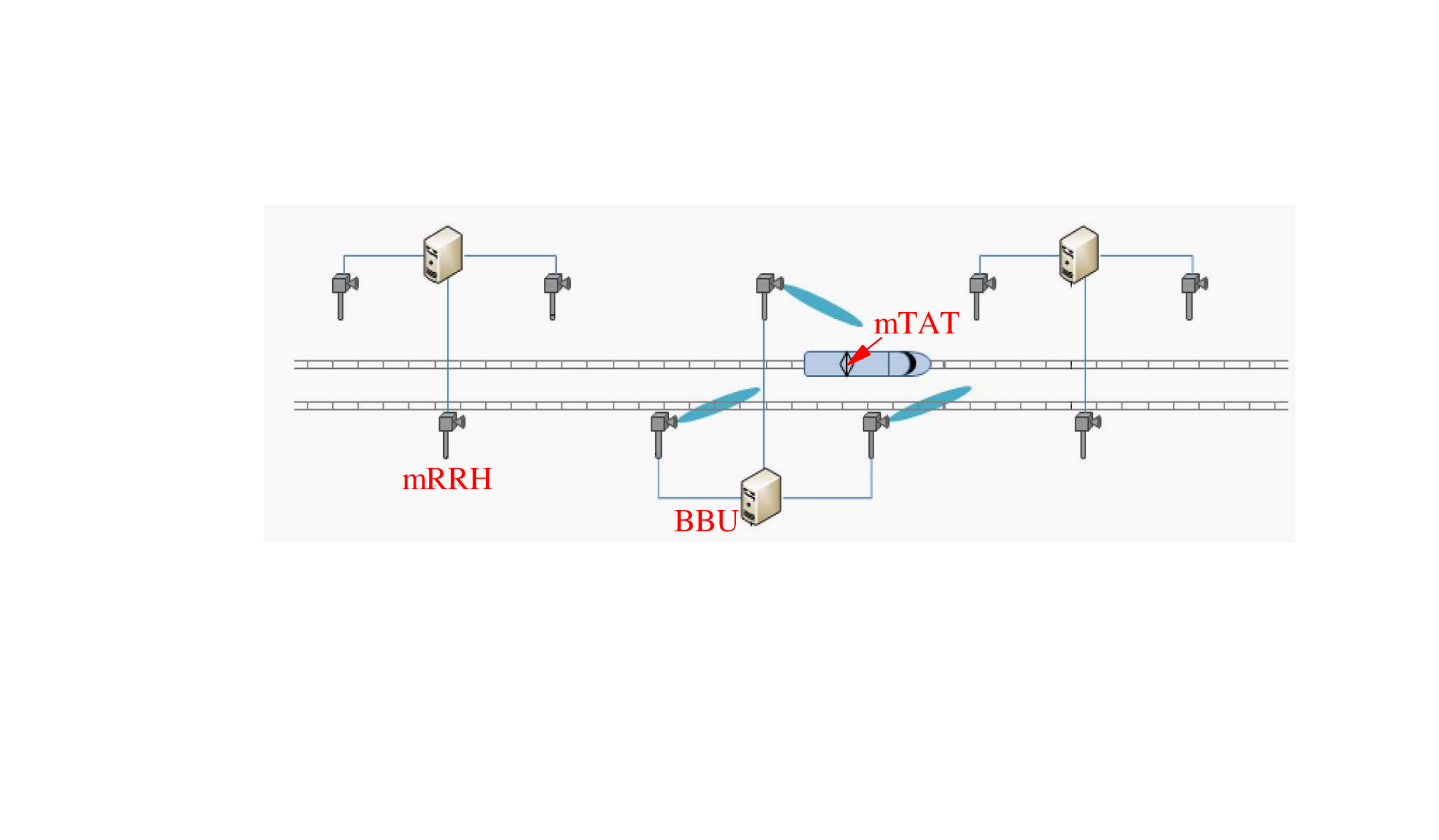}
\caption{The mmWave HST communication system.
\label{HSTfig1}}
\end{figure}

A proper network configuration is the basis of broadband wireless communications for HSTs. Refer to the third generation project partnership (3GPP) mmWave HST scenario \cite{HST07}, a potential network configuration for mmWave HST Systems can be depicted in Fig.~\ref{HSTfig1}. Along the track line, many catenary masts are placed to hang the overhead line for supplying the electricity energy of HSTs. Every few hundred meters, an mmWave remote radio head (mRRH) is installed on a catenary mast to provide wireless connections for the HSTs. Several nearby mRRHs are further connected to a common baseband processing unit (BBU) via high-bandwidth, low-latency optical transport network \cite{HST36}. The baseband processing is centralized and shared among mRRHs in the BBU. All BBUs are further connected to the core network.

Due to high penetration losses of the Faraday cage characteristics of train, direct connections between users on the fast moving HST and mRRHs are considered to be extremely difficult. Therefore, an mmWave train access terminal (mTAT) is adopted to deal with the rapid channel variation and frequent handovers from the users. Each mTAT is deployed on the front roof of the train and connected to an in-cabin wireless access point which owns several radio interfaces of multiple access technologies such as LTE and WiFi. The users in the train access wireless services via these in-cabin access points. The on-roof mTAT communicates with mRRHs, and relays the data traffic between the users and the broadband wireless networks. Both mRRHs and mTATs employ the mmWave beamforming techniques to further increase the signal quality. Accordingly, the mmWave HST communication is divided into the backhaul link between mRRHs and mTAT outsides the trains, and the access link between in-cabin wireless access points and users within the trains. It should be noted that the scenario of access link is similar to typical cellular communications which has been discussed extensively for many years. Therefore, the main challenge for mmWave HST communications lies in the backhaul link and will be studied in this paper.

\subsection{Sparse mmWave channel model in HST environments}
\label{section2b}

It is known that mmWave channel will not likely follow the rich scattering model assumed at low frequencies due to the limited number of scatters in the mmWave propagation environments. The limited multipath components have different physical transmit steering angles and receive steering angles, i.e., physical angles of departure (AoDs) and angles of arrival (AoAs). For simplicity, each scatter is assumed to contribute a single propagation path. So far, the geometric Saleh-Valenzuela channel model has been often adopted to characterize the low rank and spatial correlation of mmWave channels in static or low mobility environments \cite{HST10}. However, in rapidly time-varying HST environments, the effects of Doppler shift must be considered. Without loss of generality, it is assumed that the uniform linear antenna arrays (ULAs) are deployed at both mRRH and mTAT. Mathematically, the mmWave channel model of the backhaul link can be expressed as \cite{HST34}
\begin{equation}
{\bf{H}} = \sqrt {{N_{\rm{t}}}{N_{\rm{r}}}} \sum\limits_{l = 1}^L {{\alpha _l}{\bf{a}}\left( {{N_{\rm{r}}},\phi _l^{\rm{r}}} \right){{\bf{a}}^{\rm{H}}}\left( {{N_{\rm{t}}},\phi _l^{\rm{t}}} \right)} {e^{j2\pi {\nu _l}t}}
\label{HSTeq01}
\end{equation}
where $N_{\rm{t}}$ is the element number of ULA at mRRH, $N_{\rm{r}}$  is the element number of ULA at mTAT, $L$ is the number of effective channel paths corresponding to the limited number of scatters, ${\alpha_l} \in \mathbb{C}$ is the complex gain associated with the $l^{\rm{th}}$ path, $\phi _l^{\rm{r}}$ and $\phi _l^{\rm{t}}$ are the AoAs and AoDs, respectively, and $\nu_l$ denotes the Doppler shift associated with the $l^{\rm{th}}$ path. It can be known from \cite{HST22} that the line-of-sight (LoS) path always exists in rapidly time-varying HST environments which is quite different from typical urban environments. Without loss of generality, the first path, i.e., $l=1$, is the LoS path and others are the non-line-of-sight (NLoS)/reflected paths. In (\ref{HSTeq01}), ${\bf{a}}\left( {{N_{\rm{r}}},\phi _l^{\rm{r}}} \right) \in {{\mathbb{C}}^{{N_{\rm{r}}}}}$ and ${\bf{a}}\left( {{N_{\rm{t}}},\phi _l^{\rm{t}}} \right) \in {{\mathbb{C}}^{{N_{\rm{t}}}}}$ are the array response vectors associated with the mTAT and mRRH, respectively, and depend on the antenna array structures. For the ULA with $N$ elements, the array steering vector can be presented as \cite{HST33}
\begin{equation}
{{\bf{a}}_{{\rm{ULA}}}}\left( {N,\phi } \right){\rm{ = }}\frac{1}{{\sqrt N }}{\left[ {1,{e^{j\frac{{2\pi }}{\lambda }d\sin \phi }}, \cdots ,{e^{j\left( {U - 1} \right)\frac{{2\pi }}{\lambda }d\sin \phi }}} \right]^{\rm{T}}}
\label{HSTeq02}
\end{equation}
where $\lambda$ is the carrier wavelength and $d$ is the inter-element spacing. Note that here we omit the superscripts $\{\rm{r},\rm{t} \}$ in (\ref{HSTeq01}).

\begin{figure}[ht]
\centering
\includegraphics[width=4in]{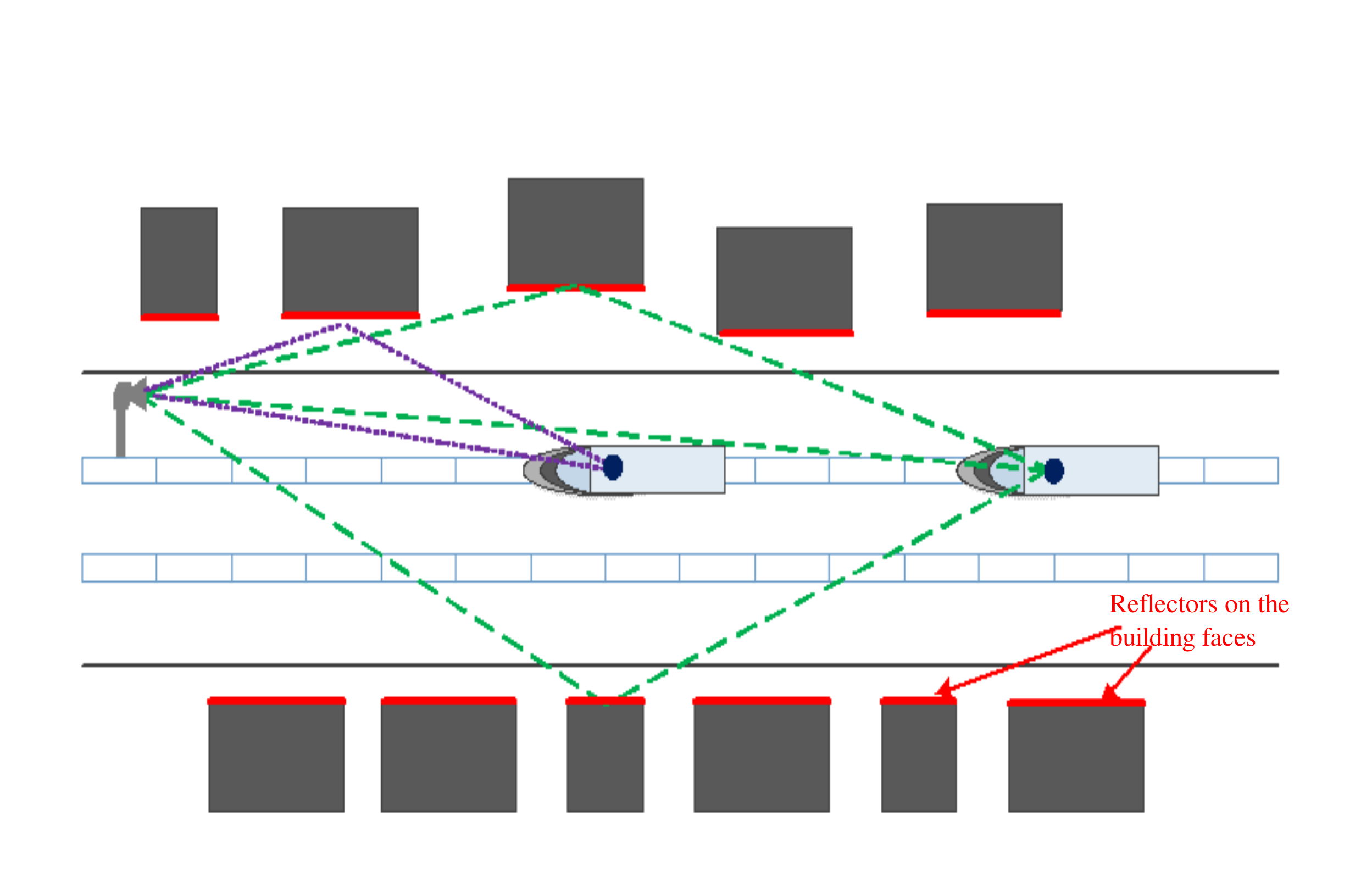}
\caption{Top view of the mmWave HST scenario with reflected paths.
\label{HSTfigB}}
\end{figure}

To design an efficient beam searching scheme, the channel dynamics must be taken into consideration. Fig.~\ref{HSTfigB} illustrate a simplified model of typical propagation environments in mmWave HST scenario, which shows dynamic multipath propagation environment. It can be seen that there exist many reflectors on building faces. Each reflector can generate a NLoS/reflected path for the moving HST when it arrives at a specific location. Since the HST is moving at a high speed, all NLoS/reflected paths can only exist for a very short time. Moreover, the generation or disappearance of a  NLoS/reflected path is difficult to predict in advance. Mathematically, except for the LoS path, the lifetime of the NLoS/reflected paths can be characterized by the birth and death process \cite{HST22}. In other words, the number of effective channel paths $L$ in (\ref{HSTeq01}) is no less than $1$ and changes over time. In order to describe the lifetime of NLoS paths, the system time can be divided into consecutive wide-sense stationary (WSS) time windows $\Delta W$. Usually one $\Delta W$ consists of multiple TTIs. For a WSS window, whether there is a newly emerging path is determined by the path number in previous WSS window, which can be described by a conditional binary distribution. If there is a new path generated in the $w_1^{\rm{th}}$ WSS window, the terminated WSS window of the new path can be expressed as
\begin{equation}
{w_2} = {w_1} + \frac{T}{{\Delta W}}
\label{HSTeq03}
\end{equation}
where $T$ is the lifetime of new path and follows a truncated Gaussian distribution. The behaviors of each path, including complex gain and Doppler shift, change over time.

\subsection{Signal model}
\label{section2c}

\begin{figure}[ht]
\centering
\subfigure[$N=32$]{\includegraphics[width=3.2in]{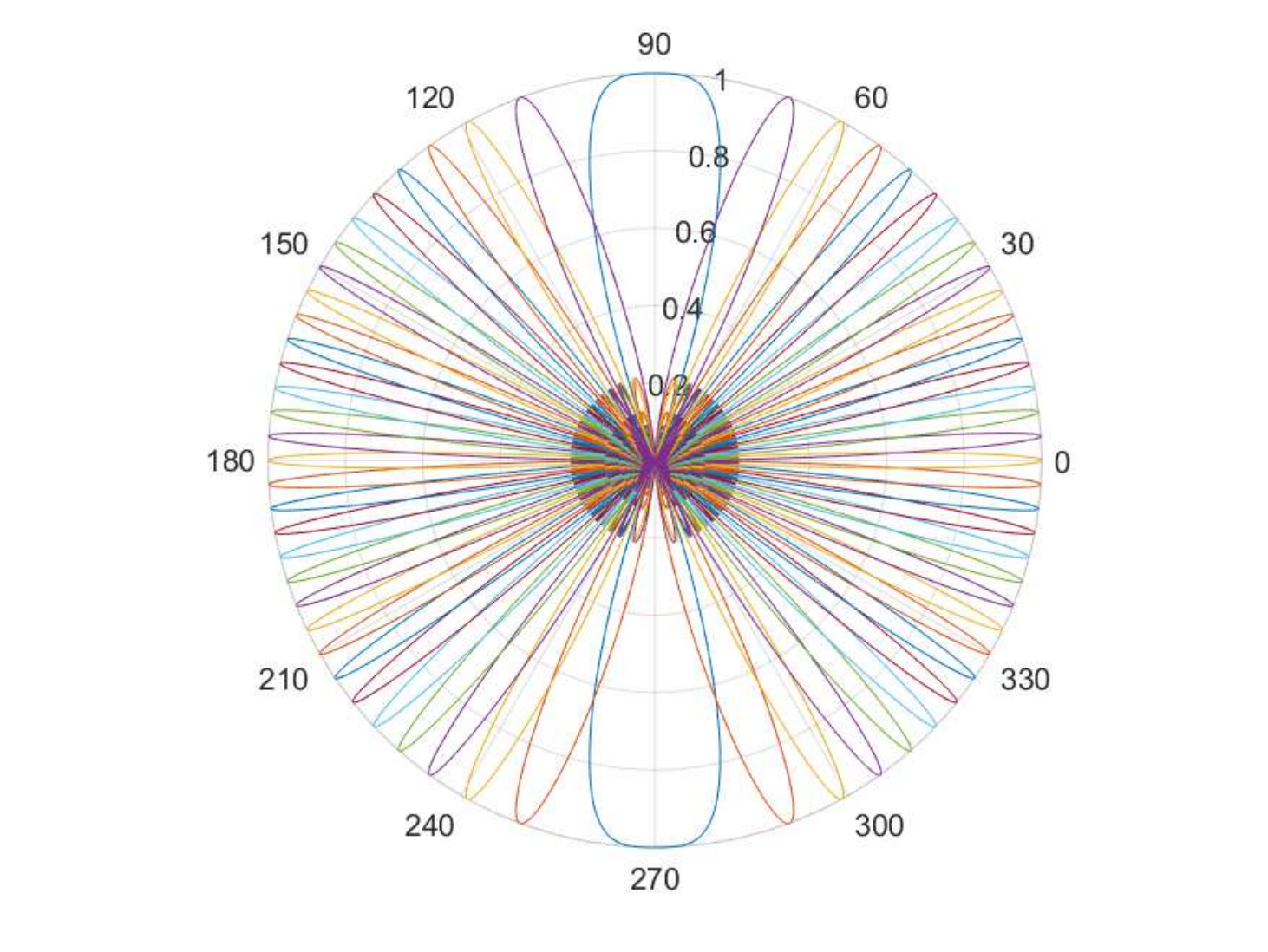}}
\subfigure[$N=64$]{\includegraphics[width=3.2in]{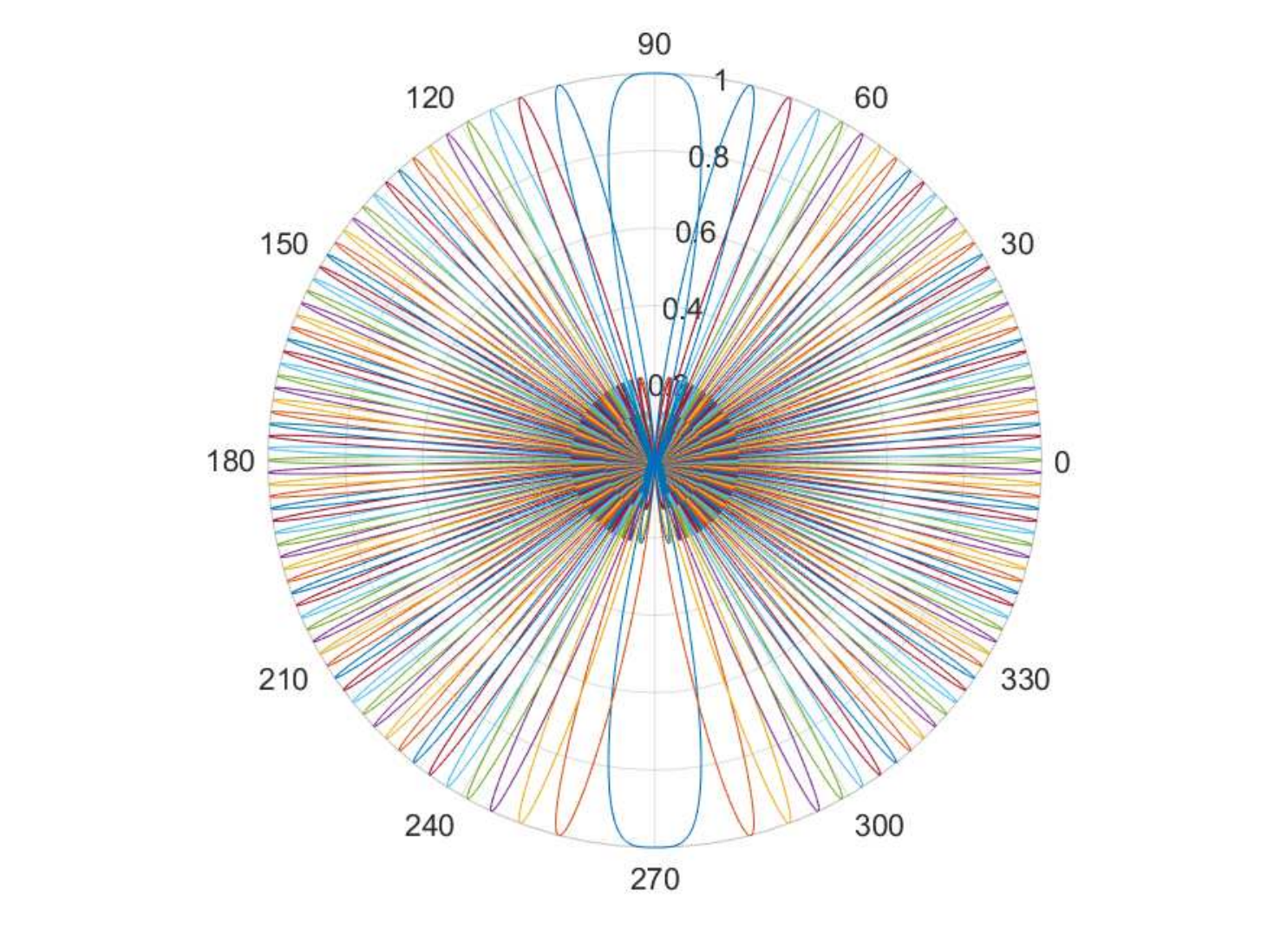}}
\caption{Beam patterns of antenna array.\label{HSTfig2}}
\end{figure}

In order to measure the mmWave channel, the mRRH broadcasts a sequence of beamformed pilot signals to mTAT. The transmitting and receiving beamforming operations are often realized by the networks of RF phase shifters. Let ${{\bf{f}}_p} \in {\mathbb{C}^{{N_{\rm{t}}} \times 1}}$ and ${{\bf{w}}_q} \in {\mathbb{C}^{{N_{\rm{r}}} \times 1}}$  be the transmitting and receiving beamforming vectors, respectively. Then, all elements of ${{\bf{f}}_p}$ and ${{\bf{w}}_p}$ have constant modulus and unit norm, i.e., $\left\| {{{\bf{f}}_p}} \right\|{\rm{ = }}1$, $\forall p$, and $\left\| {{{\bf{w}}_q}} \right\|{\rm{ = }}1$, $\forall q$. Considering the hardware constraints, each phase shifter is digitally controlled and can only choose one column from a discrete Fourier transformation (DFT) matrix \cite{HST24}. The DFT matrix of mRRH (mTAT) consists of the array steering vectors of $N_{\rm{t}}$ ($N_{\rm{r}}$) orthogonal beams spread over the entire angular domain, i.e.,
\begin{equation}
{{\bf{A}}_{\rm{t}}} = {\left[ {{\bf{a}}\left( {{N_{\rm{t}}},\arcsin \left( {{\theta _1}} \right)} \right), \cdots ,{\bf{a}}\left( {{N_{\rm{t}}},\arcsin \left( {{\theta _{{N_{\rm{t}}}}}} \right)} \right)} \right]^{\rm{H}}}
\label{HSTeq04}
\end{equation}
\begin{equation}
{{\bf{A}}_{\rm{r}}} = {\left[ {{\bf{a}}\left( {{N_{\rm{r}}},\arcsin \left( {{\varphi _1}} \right)} \right), \cdots ,{\bf{a}}\left( {{N_{\rm{r}}},\arcsin \left( {{\varphi _{{N_{\rm{r}}}}}} \right)} \right)} \right]^{\rm{H}}}
\label{HSTeq05}
\end{equation}
where ${\theta _n} = \left( n-\left(N_{\rm{t}}+1\right)/2\right)/N_{\rm{t}}$ with $n = 1, \cdots ,{N_{\rm{t}}}$  and ${\phi _n} = \left( n-\left(N_{\rm{r}}+1\right)/2\right)/N_{\rm{r}}$ with $n = 1, \cdots ,{N_{\rm{r}}}$ are the normalized spatial directions for mRRH and mTAT, respectively. Fig.~\ref{HSTfig2} shows the normalized beamforming patterns when $N=32$ and $N=64$.

During the channel estimation, if the $p^{\rm{th}}$ transmitting and the $q^{\rm{th}}$ receiving beamforming vectors are adopted at mRRH and mTAT, respectively, the received signals can be described as
\begin{equation}
{{\bf{y}}_{q,p}} = \sqrt P {\bf{w}}_q^{\rm{H}}{\bf{H}}{{\bf{f}}_p}{{\bf{s}}_p} + {\bf{w}}_q^{\rm{H}}{{\bf{n}}_{q,p}}
\label{HSTeq06}
\end{equation}
where $P$ is the transmit power within the channel estimation period, ${\bf{s}}_p$ is the pilot sequence on the $p^{\rm{th}}$ transmitting beam, and ${{\bf{n}}_{q,p}} \sim {\cal{CN}}\left( {0,{\sigma ^2}{{\bf{I}}_{{N_{\rm{r}}}}}} \right)$ is the complex Gaussian noise vector corrupting the receiving signals. Note that the transmitted signal $\bf{s}$ in (\ref{HSTeq06}) satisfies
\begin{equation}
{\rm{E}}\left\{ {{\bf{s}}{{\bf{s}}^{\rm{H}}}} \right\} = {{\bf{I}}_{{N_{\rm{S}}}}}
\label{HSTeq07}
\end{equation}

\section{Beam searching problem of mmWave HST communications}
\label{section3}

Using (\ref{HSTeq04}) and (\ref{HSTeq05}) to measure the mmWave channel, the angle quantization errors are too small to be noticed if the antenna array size is large enough. Accordingly, the physical channel matrix $\bf{H}$ can be mapped to a virtual channel matrix (VCM) $\bf{H}_{\rm{V}}$ through the following relationship
\begin{equation}
{{\bf{H}}_{\rm{V}}} = \frac{1}{{\sqrt {{N_{\rm{t}}}{N_{\rm{r}}}} }}{{\bf{A}}_{\rm{r}}}^{\rm{H}}{\bf{H}}{{\bf{A}}_{\rm{t}}}
\label{HSTeq08}
\end{equation}
The $(i,j)$$^{\rm{th}}$ element of $\bf{H}_{\rm{V}}$ represents the channel state when the virtual angles seen by mRRH and mTAT are $\theta_i$ and $\phi_j$, respectively. For the entry of $\bf{H}_{\rm{V}}$ corresponding to the $l$$^{\rm{th}}$ path, the value is given by ${\alpha _l}\exp \left( {j2\pi {\nu _l}t} \right)$. The sparse property of the physical channel model (\ref{HSTeq01}) results in that there are $L$ non-zero elements among $N_{\rm{r}} N_{\rm{t}}$ entries of the VCM $\bf{H}_{\rm{V}}$. In practice, the value of $L$ changes with the location of HST. Therefore, searching for available paths in dynamic environments can be formulated as a beam searching problem which tries to find Tx-Rx beam pairs $(i,j)$ with highest measured path gains for data transmissions.

In order to tackle the huge path loss in mmWave channels, the massive antenna arrays are deployed at both mRRH and mTAT. Due to the large antenna array size, it would be time-consuming to make a full estimation of mmWave channel by exploring all possible beams. Especially in HST scenarios, the channel varies quickly over time which requires that the channel estimation must be carried out frequently. Moreover, the period of each TTI is too short to allocate enough time for accurate channel estimation. Intuitively, the sparsity property of mmWave channels can be exploited to only search and estimate the several strong channel paths with low overhead. It should be noted that the AoD and AoA associated with all paths are changed with the location of HST. Moreover, due to the installation errors of antenna arrays, even the AoD and AoA associated with the always existing LoS path cannot be predicted before the system operation. However, since the beamwidth of large antenna array is very narrow, a small variation of AoD or AoA associated with each path will result in a significant fluctuation of beamforming gain. Therefore, the key challenge is how to design an efficient beam searching scheme in such a way that the strong channel paths can be quickly searched, which will leave more time for data transmission to achieve a higher throughput in practical systems.

\section{Proposed beam searching scheme}
\label{section4}

In this section, we intend to exploit the historic information to accelerate the beam searching. The beam searching problem of mmWave HST communications can be formulated as a restless MAB. Following the bandit terminology, each entry in $\bf{H}_{\rm{V}}$ is named as an arm. Therefore, there are $N_{\rm{r}} N_{\rm{t}}$ arms. At each timeslot $t \in \left\{ 1,2, \cdots \right\}$ (corresponding to the period of one TTI), totally $M$ arms are selected from the $N_{\rm{r}} N_{\rm{t}}$ arms for channel estimation. Considering the limitation of mmWave HST communications, $M$ is far less than $N_{\rm{r}} N_{\rm{t}}$. After channel estimation, $D$ ($D < M$ ) arms with $D$ largest path gains are chosen for data transmission. Each chosen arm $i \in \left\{ {1, \cdots ,{N_{\rm{t}}}{N_{\rm{r}}}} \right\}$ contributes a particular transmission rate, here and thereafter known as reward ${x_i}\left( t \right)$. The rewards ${x_i}\left( t \right)$ are calculated according to the received signal and this information is fed back to update the information of the beaming searching.

The well-known upper confidence bound (UCB) algorithms proposed in \cite{HST25} and their variants are widely applied to many bandit problems. The basic idea of UCB algorithms is to estimate the unknown expected reward of each arm by making a linear combination of previously observed rewards of the arms. Inspired by basic idea of UCB algorithms, this section will propose one scheme to deal with beam searching problem in mmWave HST communications. To facilitate the understanding, the detailed procedure are described as follows.

During the whole procedure, the proposed scheme maintains two kinds of bandit information. The first one is the number of times that arm $i$ has been chosen up to timeslot $t$, denoted as $n_i$, and the second one is the average of historical rewards obtained for arm $i$, namely $\mu_i$. All bandit information will be initialized to zero.

At the beginning of each timeslot, mRRH will search all available paths to maximize the transmission rate. Although historical information can be exploited to shorten the beam searching, the practical propagation environments of HST communications are not static. For example, a growing tree besides the track line may create or eliminate some channel paths. In other words, the expected reward of associated arm varies very slowly with time. Therefore, the exploration is always needed. Since $\mu_i$ is regarded as an estimate of the true expected reward of arm $i$, the arms which own the largest UCBs of rewards are chosen. Formally, the UCB of arm  $i$ is calculated as
\begin{equation}
A\left( i \right) = {\mu _i} + c\sqrt {\frac{{\ln t}}{{{n_i}}}} \quad \forall i \in \left\{ {1, \cdots ,{N_{\rm{t}}}{N_{\rm{r}}}} \right\}
\label{HSTeq09}
\end{equation}
where $c$ is a constant that determines the width of the confidence bound to control degree of exploration. Using (\ref{HSTeq09}), $M$ arms with largest UCBs can be selected such that
\begin{equation}
A\left( {{j_1}} \right) \ge A\left( {{j_2}} \right) \ge  \cdots  \ge A\left( {{j_M}} \right) \ge A\left( i \right)\quad \forall i \notin \left\{ {{j_1},{j_2}, \cdots ,{j_M}} \right\}
\label{HSTeq10}
\end{equation}
It can be observed from (\ref{HSTeq09}) that if the arm with large $c\sqrt {{\ln t}/{n_i}} $ is chosen, an explorative decision is made, since in such a case taking $\mu_i$ as the estimate of the true expected reward is quite unreliable. Contrarily, if an arm with large $\mu_i$ is chosen, an exploitative decision is made. Considering that $c\sqrt {{\ln t}/{n_i}} $ decreases rapidly with time, the number of explorative decisions is limited. Moreover, as $c\sqrt {{\ln t}/{n_i}} $ becomes smaller, the average $\mu_i$ gets closer to the true expected reward. Accordingly, the arms with the largest expected rewards are associated with the most promising paths that can be exploited to maximize the transmission rate.

After the arm selection, an arm set can be obtained as
\begin{equation}
{G_M} \buildrel \Delta \over = \left\{ {{j_1},{j_2}, \cdots ,{j_M}} \right\}
\label{HSTeq11}
\end{equation}
For any arm $i$ in $G_M$, the channel state of the associated path is the $\left( {q\left( i \right),p\left( i \right)} \right)$$^{\rm{th}}$ element of ${{\bf{H}}_{\rm{V}}}$. Therefore, the $p(i)$$^{\rm{th}}$ transmitting and $q(j)$$^{\rm{th}}$ receiving beamforming vectors will be used by mRRH and mTAT, respectively, to measure the associated path of arm $i$. Then, the mTAT receives the signals as
\begin{equation}
{{\bf{y}}_{q\left( i \right),p\left( i \right)}} = \sqrt P {\bf{w}}_{q\left( i \right)}^{\rm{H}}{\bf{H}}{{\bf{f}}_{p\left( i \right)}}{{\bf{s}}_{p\left( i \right)}} + {\bf{w}}_{q\left( i \right)}^{\rm{H}}{{\bf{n}}_{q\left( i \right),p\left( i \right)}}\quad i \in {G_M}
\label{HSTeq12}
\end{equation}
Then the mean received pilot signal power is obtained as
\begin{equation}
Y\left( i \right) = \frac{1}{{{N_{\rm{p}}}}}\left\| {{{\bf{y}}_{q\left( i \right),p\left( i \right)}}} \right\|_2^2\quad i \in {G_M}
\label{HSTeq13}
\end{equation}
where $N_{\rm{p}}$ is the length of pilot sequence. Meanwhile, the information of channel state, including complex gains and Doppler shifts, can be obtained by classic estimation techniques, which is beyond the scope of this paper. Among the measured $M$ paths, only $D$ paths with best channel conditions will be chosen for further data transmissions. Mathematically, the associated arms of the chosen $D$ paths, namely $k_1$, $k_2$, $\cdots$, and $k_D$, are taken from $G$ such that
\begin{equation}
Y\left( {{k_1}} \right) \ge Y\left( {{k_2}} \right) \ge  \cdots  \ge Y\left( {{k_D}} \right) \ge Y\left( i \right)\quad \forall i \in {G_M}\backslash \left\{ {{k_1},{k_2}, \cdots ,{k_D}} \right\}
\label{HSTeq14}
\end{equation}
To facilitate the following description, let ${G_D} \buildrel \Delta \over = \left\{ {{k_1},{k_2}, \cdots ,{k_D}} \right\}$ be the set of paths that is used for data transmissions.

During the data transmission period, parallel $D$ data streams will be transmitted over the chosen $D$ paths. At the same time, the mTAT will measure the transmission rate over each path. The measured rates will be fed back to the mRRH to calculate the rewards of the associated arms. The reward of a chosen arm is defined as a utility of its measured rate. Finally, the bandit information of all unchosen arms can be updated as
\begin{equation}
{\mu _i} = \frac{{{x_i} + {\mu _i}{n_i}}}{{{n_i} + 1}}\;\quad \forall i \in {G_D}
\label{HSTeq15}
\end{equation}
\begin{equation}
{n_i} = {n_i} + 1\;\quad \forall i \in {G_D}
\label{HSTeq16}
\end{equation}

The proposed beam searching scheme is shown as Table~\ref{HSTtab1}.
\begin{table}[ht]
\caption{{Algorithm 1 Proposed bandit inspired beam searching algorithm.}}
\label{HSTtab1}
\begin{center}
\begin{tabular}{l}
\hline\hline
1. $t=0$, $n_i=0$, $\mu_i=0$, $\forall i \in \left\{ {1, \cdots ,{N_{\rm{R}}}{N_{\rm{T}}}} \right\}$; {\quad}// Initialization\\
2. While 1 do {\quad}// Mainloop\\
3. {\qquad}$t=t+1$;\\
4. {\qquad}Use (\ref{HSTeq09}) to calculate the UCBs of all arms and update $G_M$ to satisfy (\ref{HSTeq10});\\
5. {\qquad}Measure the channel state over the associated paths in $G_M$;\\
6. {\qquad}Choose $D$ paths using (\ref{HSTeq14}) for data transmissions;\\
7. {\qquad}Feedback the transmission rates of the chosen $D$ paths as the rewards of the associated arms;\\
8. {\qquad}Update the bandit information using (\ref{HSTeq15}) and (\ref{HSTeq16}).\\
9. End while\\
\hline\hline
%\centering Initialization
%\tabularnewline\hline
%\centering $t = 0;{n_i} = 0,{\mu _i} = 0\;\;\forall i \in \left\{ {1, \cdots ,{N_{\rm{R}}}{N_{\rm{T}}}} \right\}$
%\tabularnewline\hline\hline
\end{tabular}
\end{center}
\end{table}

\section{Regret analysis}
\label{section5}

In MAB, a quantity termed as expected cumulative regret is often used to characterize the learning performance, which represents the cumulative difference between the reward of the chosen arms and the maximum expected reward, which is attainable by a "genie" who knows the expected reward of all arms. Accordingly, the objective of beam searching is equivalent to minimizing the expected cumulative regret. The expected cumulative regret can be expressed as
\begin{equation}
{R_T} = T\sum\limits_{i \in {G_D}} {{\mu _i}}  - {\rm{E}}\left\{ {\sum\limits_{t = 1}^T {\sum\limits_{i = 1}^D {{x_i}\left( t \right){I_i}\left( t \right)} } } \right\}
\label{HSTeq17}
\end{equation}
where ${I_i}\left( t \right)$ is $1$ at timeslot  $t$ if arm $i$ is selected and $0$ otherwise. This section will derive an upper bound of the regret of the proposed beam scheme in (\ref{HSTeq17}).

\begin{Lemma}
\label{HSTLemma1}
Under the proposed algorithm, the expected number of times that any arm $i \notin G_D$ is chosen after $n$ timeslots satisfy
\begin{equation}
{\rm{E}}\left\{ {{T_i}\left( n \right)} \right\} \le \frac{{4{c^2}\ln n}}{{\Delta _{\min }^2}} + 1 + \frac{{{\pi ^2}}}{3}
\label{HSTeq18}
\end{equation}
where ${T_i}\left( n \right)$ is the number of times that arm $i$ is chosen after $n$ timeslots, ${\Delta _{\min }}$ is the minimum expected reward between arm $i$ and any arm in $G_D$, i.e.,
\begin{equation}
{\Delta _{\min }} = \mathop {\min }\limits_{j \in {G_D}} \left| {{\mu _j} - {\mu _i}} \right|
\label{HSTeq19}
\end{equation}
\begin{IEEEproof}
To simplify the proof, we assume that there does not exist two arms with the same expected rewards. Mathematically, ${T_i}\left( n \right)$ can be expressed as
\begin{eqnarray}
{T_i}\left( n \right) &=& \sum\limits_{t = {\rm{1}}}^n {{\rm{I}}\left\{ {{\rm{arm~}}i{\rm{~is~chosen~at~ TTI~}}t{\rm{ }}} \right\}} \nonumber\\
 &\le& l + \sum\limits_{t = {\rm{1}}}^n {{\rm{I}}\left\{ {{\rm{arm~}}i{\rm{~is~chosen~at~TTI~}}t,\;{T_i}\left( {t - 1} \right) \ge l{\rm{ }}} \right\}}
\label{HSTeq20}
\end{eqnarray}
where ${\rm{I}}\left\{ x \right\}$ is the indicator function defined to be $1$ when the predicate $x$ is true, and $0$ when it is false. Note that if one arm $i \notin {G_D}$ is chosen at timeslot $t$, there must exist an arm $j\left( t \right) \in {G_D}$ satisfying the following inequality
\begin{equation}
{\hat \mu _{i,{T_i}\left( {t - 1} \right)}} + {C_{t - 1,{T_i}\left( {t - 1} \right)}} \ge {\hat \mu _{j\left( t \right),{T_{j\left( t \right)}}\left( {t - 1} \right)}} + {C_{t - 1,{T_{j\left( t \right)}}\left( {t - 1} \right)}}
\label{HSTeq21}
\end{equation}
where ${\hat \mu _{i,\;{n_i}}}$ is the mean of all the observed rewards of the arm $i$ when it is observed $n_i$ times, and ${C_{t,s}} = c\sqrt {{\ln t}/s} $.  Since the arm $i$ does not belong to $G_D$, it can be concluded that
\begin{equation}
{\mu _i} < {\mu _{j\left( t \right)}}
\label{HSTeq22}
\end{equation}
Therefore, we have
\begin{eqnarray}
&&\sum\limits_{t = {\rm{1}}}^n {{\rm{I}}\left\{ {{\rm{arm~}}i{\rm{~is~chosen~at~TTI~}}t,{T_i}\left( {t - 1} \right) \ge l{\rm{ }}} \right\}} \nonumber\\
 &\le& \sum\limits_{t = {\rm{1}}}^n {{\rm{I}}\left\{ {{{\hat \mu }_{i,\;{T_i}\left( {t - 1} \right)}} + {C_{t - 1,{T_i}\left( {t - 1} \right)}} \ge {{\hat \mu }_{j\left( t \right),\;{T_{j\left( t \right)}}\left( {t - 1} \right)}} + {C_{t - 1,{T_{j\left( t \right)}}\left( {t - 1} \right)}},\;{T_i}\left( {t - 1} \right) \ge l{\rm{ }}} \right\}} \nonumber\\
 &\le& \sum\limits_{t = {\rm{1}}}^n {{\rm{I}}\left\{ {\mathop {\max }\limits_{l \le {n_i} < t} {{\hat \mu }_{i,\;{n_i}}} + {C_{t - 1,{n_i}}} \ge \mathop {\min }\limits_{0 \le {n_{j\left( t \right)}} < t} {{\hat \mu }_{j\left( t \right),{n_{j\left( t \right)}}}} + {C_{t - 1,{n_{j\left( t \right)}}}}{\rm{ }}} \right\}} \nonumber\\
 &\le& \sum\limits_{t = 1}^\infty  {\sum\limits_{{n_{j\left( t \right)}} = 0}^{t - 1} {\sum\limits_{{n_i} = l}^{t - 1} {{\rm{I}}\left\{ {{{\hat \mu }_{i,\;{n_i}}} + {C_{t,{n_i}}} \ge {{\hat \mu }_{j\left( t \right),{n_{j\left( t \right)}}}} + {C_{t,{n_{j\left( t \right)}}}}{\rm{ }}} \right\}} } }
\label{HSTeq23}
\end{eqnarray}
The inequality ${\hat \mu _{i,\;{n_i}}} + {C_{t,{n_i}}} \ge {\hat \mu _{j\left( t \right),{n_{j\left( t \right)}}}} + {C_{t,{n_{j\left( t \right)}}}}$indicates that at least one of following inequalities must be held
\begin{equation}
{\hat \mu _{i,\;{n_i}}} \ge {\mu _i} + {C_{t,{n_i}}}
\label{HSTeq24}
\end{equation}

\begin{equation}
{\hat \mu _{j\left( t \right),{n_{j\left( t \right)}}}} \le {\mu _{j\left( t \right)}} - {C_{t,{n_{j\left( t \right)}}}}
\label{HSTeq25}
\end{equation}

\begin{equation}
{\mu _i} + 2{C_{t,{n_i}}} \ge {\mu _{j\left( t \right)}}
\label{HSTeq26}
\end{equation}
Using the Chernoff-Hoeffding bound \cite{HST25}, the upper bounds of the probabilities that the two inequalities (\ref{HSTeq24}) and (\ref{HSTeq25}) hold are
\begin{equation}
\Pr \left\{ {{{\hat \mu }_{i,\;{n_i}}} \ge {\mu _i} + {C_{t,{n_i}}}} \right\} \le {e^{ - 4\ln t}} = {t^{ - 4}}
\label{HSTeq27}
\end{equation}

\begin{equation}
\Pr \left\{ {{{\hat \mu }_{j\left( t \right),{n_{j\left( t \right)}}}} \le {\mu _{j\left( t \right)}} - {C_{t,{n_{j\left( t \right)}}}}} \right\} \le {e^{ - 4\ln t}} = {t^{ - 4}},
\label{HSTeq28}
\end{equation}
respectively. From (\ref{HSTeq26}), we can derive that
\begin{eqnarray}
\Pr \left\{ {{\mu _i} + 2{C_{t,{n_i}}} \ge {\mu _{j\left( t \right)}}} \right\} &=& \Pr \left\{ {{\mu _{j\left( t \right)}} - {\mu _i} - 2{C_{t,{n_i}}} \le 0} \right\}\nonumber\\
 &\le& \Pr \left\{ {\mathop {\min }\limits_{j\left( t \right) \in {G_D}} \left\{ {{\mu _{j\left( t \right)}}} \right\} - {\mu _i} - 2{C_{t,{n_i}}} \le {\rm{0}}} \right\}\nonumber\\
&=&\Pr\left\{ {{\Delta _{\min }} - 2c\sqrt {\frac{{\ln t}}{{{n_i}}}}  \le {\rm{0}}} \right\}
\label{HSTeq29}
\end{eqnarray}
By some algebra operations, we can derive that
\begin{equation}
\Pr \left\{ {{\mu _i} + 2{C_{t,{n_i}}} \ge {\mu _{j\left( t \right)}}} \right\} = 0\quad {n_i} > \frac{{4{c^2}\ln t}}{{\Delta _{\min }^2}}
\label{HSTeq30}
\end{equation}
Using (\ref{HSTeq27}), (\ref{HSTeq28}) and (\ref{HSTeq30}), the expectation of $T_i(n)$ is upper bounded by
\begin{eqnarray}
{\rm{E}}\left\{ {{T_i}\left( n \right)} \right\} &\le& \left\lceil {\frac{{4{c^2}\ln n}}{{\Delta _{\min }^2}}} \right\rceil  + \sum\limits_{t = 1}^\infty  {\sum\limits_{{n_{j\left( t \right)}} = 0}^{t - 1} {\sum\limits_{{n_i} = \left\lceil {\frac{{4{c^2}\ln n}}{{\Delta _{\min }^2}}} \right\rceil }^{t - 1} {\Pr \left\{ {{{\hat \mu }_{i,\;{n_i}}} \ge {\mu _i} + {C_{t,{n_i}}}} \right\}} } }
\nonumber\\ && + \sum\limits_{t = 1}^\infty  {\sum\limits_{{n_{j\left( t \right)}} = 0}^{t - 1} {\sum\limits_{{n_i} = \left\lceil {\frac{{4{c^2}\ln n}}{{\Delta _{\min }^2}}} \right\rceil }^{t - 1} {\Pr \left\{ {{{\hat \mu }_{j\left( t \right),{n_{j\left( t \right)}}}} \le {\mu _{j\left( t \right)}} - {C_{t,{n_{j\left( t \right)}}}}} \right\}} } } \nonumber\\
 &\le& \frac{{4{c^2}\ln n}}{{\Delta _{\min }^2}} + 1 + 2\sum\limits_{t = 1}^\infty  {\sum\limits_{{n_{j\left( t \right)}} = 0}^{t - 1} {\sum\limits_{{n_i} = 1}^{t - 1} {{t^{ - 4}}} } } \nonumber\\
 &\le& \frac{{4{c^2}\ln n}}{{\Delta _{\min }^2}} + 1 + \frac{{{\pi ^2}}}{3}
\label{HSTeq31}
\end{eqnarray}
\end{IEEEproof}
\end{Lemma}

Using Lemma~\ref{HSTLemma1}, we can derive the upper bound of the regret of our proposed algorithm.
\begin{Theorem}
\label{HSTTheorem1}
The expected regret of our proposed algorithms is upper bounded by
\begin{equation}
{R_T} \le \sum\limits_{i \notin {G_D}}^{} {{\Delta _{\max }}\left( {\frac{{4{c^2}\ln n}}{{\Delta _{\min }^2}} + 1 + \frac{{{\pi ^2}}}{3}} \right)}
\label{HSTeq32}
\end{equation}
where ${\Delta _{\min }}$ is the maximum expected reward between arm $i$ and any arm in $G_D$, i.e.,
\begin{equation}
{\Delta _{\max }} = \mathop {\max }\limits_{j \in {G_D}} \left| {{\mu _j} - {\mu _i}} \right|
\label{HSTeq33}
\end{equation}
\begin{IEEEproof}
From (\ref{HSTeq17}), we can derive that
\begin{equation}
{R_T} \le \sum\limits_{i \notin {G_D}}^{} {{\Delta _{\max }}{\rm{E}}\left\{ {{T_i}\left( n \right)} \right\}}
\label{HSTeq34}
\end{equation}
Using Lemma~\ref{HSTLemma1}, (\ref{HSTeq32}) is obtained.
\end{IEEEproof}
\end{Theorem}

Theorem~\ref{HSTTheorem1} shows the upper bound of the regret of the proposed algorithm, which grows logarithmical in time and linearly in the number of arms.

\section{Numerical results}
\label{section6}

This section shows the performance of the proposed algorithm by simulations. Refer to the 3GPP mmWave HST scenario \cite{HST07}, a simulation platform is developed by deploying several mRRHs along the track line. Without loss of generality, the distance between the neighboring mRRHs is set to be 500m and the separation distance between each RRH and side track is 5m. For the downlink transmission, the carrier frequency is set to be 28GHz. Refer to \cite{HST26}, the pathloss is modeled as
\begin{equation}
P{L_{\left[ {dB} \right]}} =  {61.4 + 34{{\lg }}\left( d \right)}
\label{HSTeq35}
\end{equation}
where $d$ is the propagation distance of transmitted signals. The antenna gains of mRRH and mTAT are assumed to be 25dBi and 12dBi, respectively. The maximum transmit power of mRRH is set to be 33dBm and the thermal noise power at the receiver of mTAT is -80dBm. In the simulations, the HST is assumed to be moving at the speed of 360km/h and the period of TTI is assumed to be 0.25ms. Accordingly, the service time of each mRRH is totally 20,000 timeslot periods.

\begin{figure}[ht]
\centering
\includegraphics[width=4in]{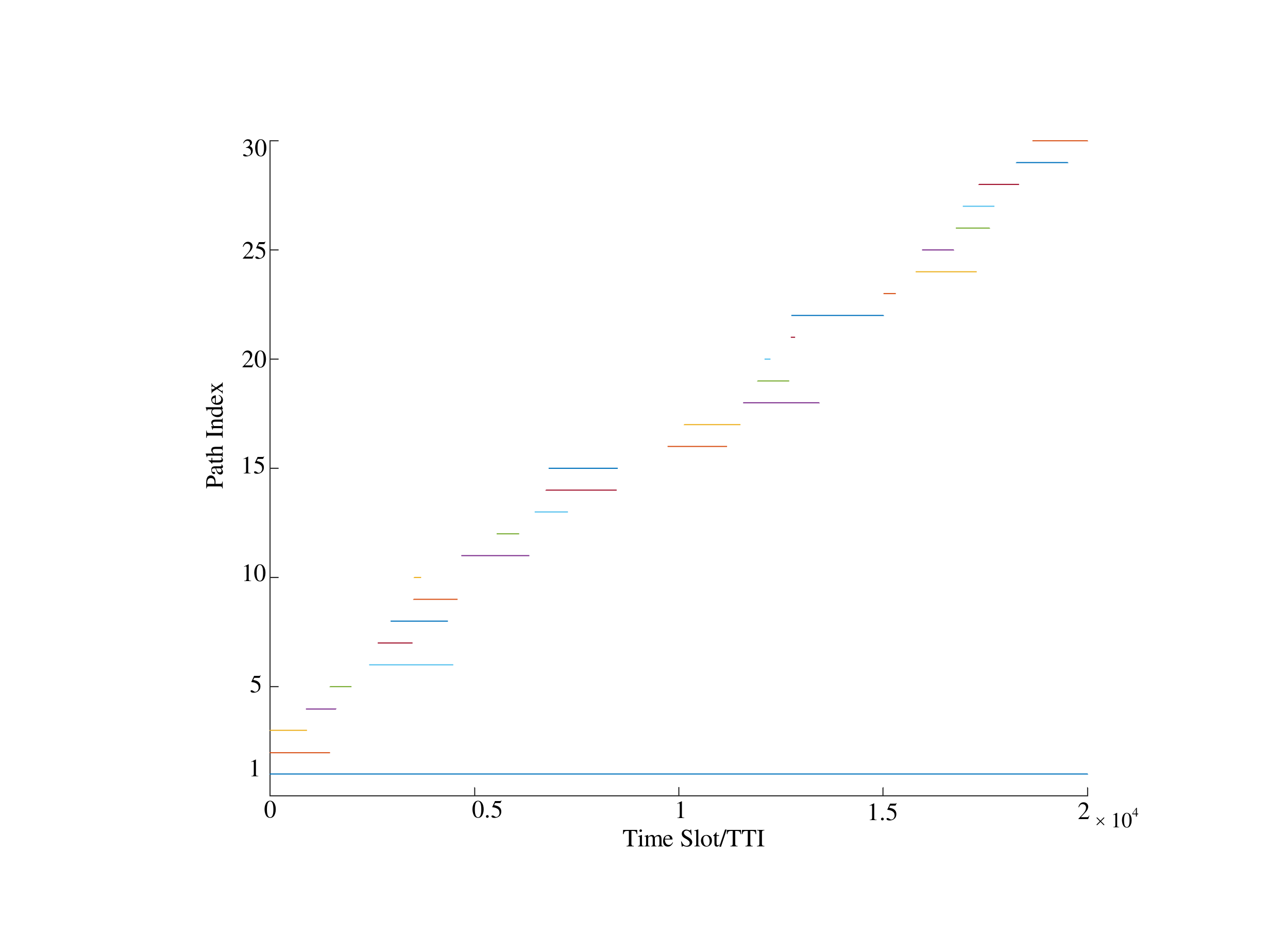}
\caption{The temporal evolution of all paths in HST scenario.
\label{HSTfigA}}
\end{figure}

\begin{figure}[ht]
\centering
\includegraphics[width=4in]{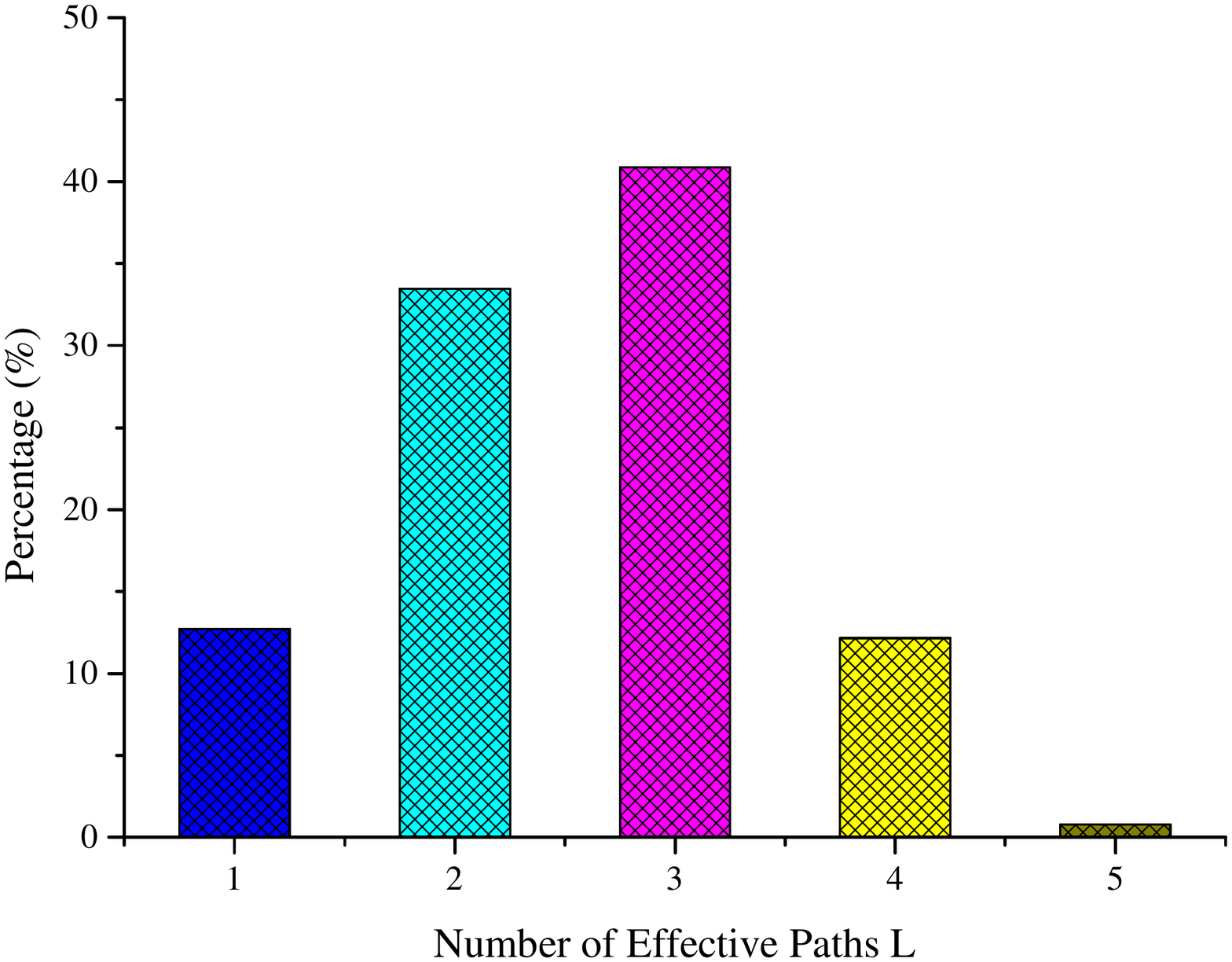}
\caption{Percentage of locations with different effective channel paths $L$ when a HST traverses the coverage of an mRRH.
\label{HSTfig3}}
\end{figure}

\begin{figure}[ht]
\centering
\includegraphics[width=4in]{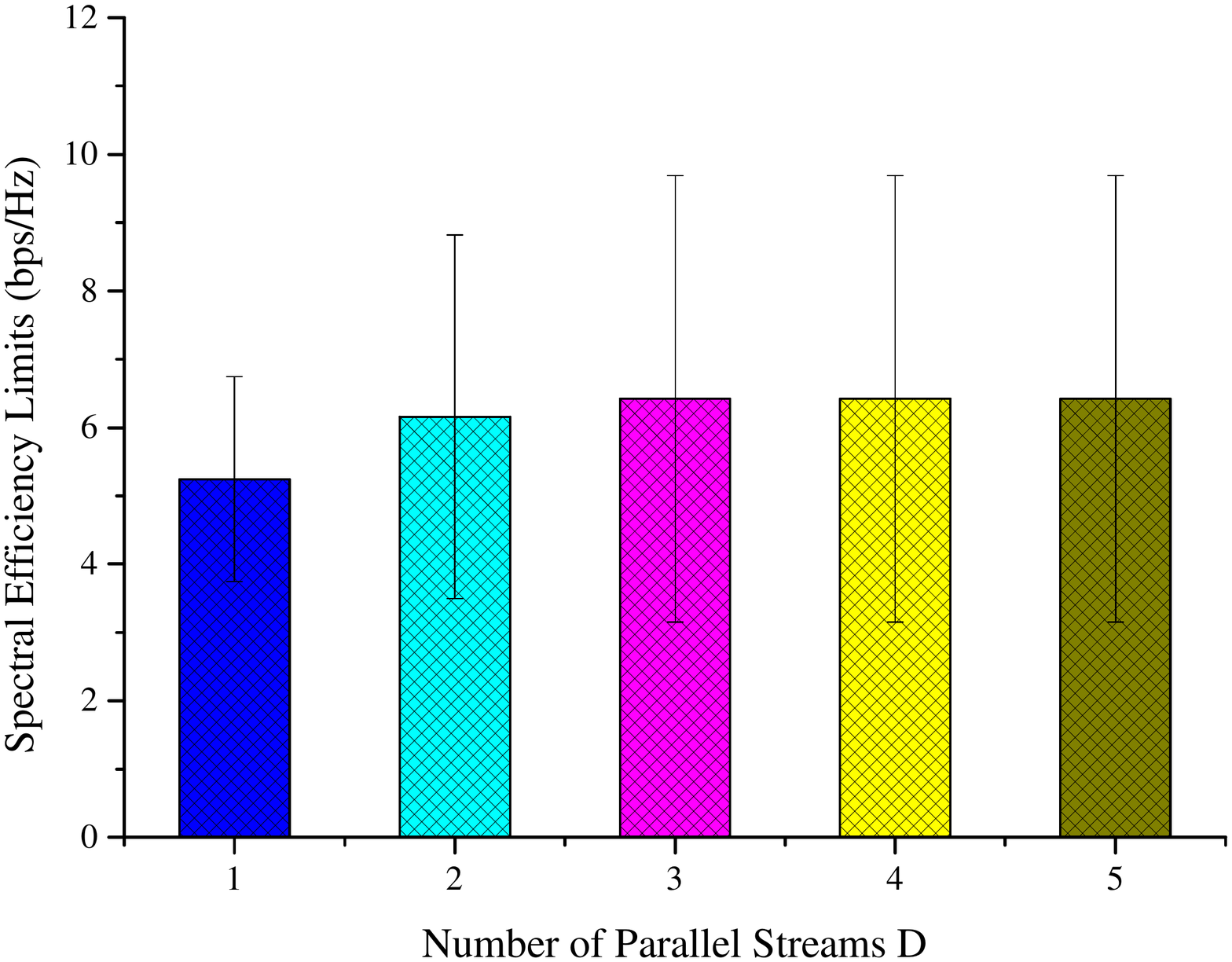}
\caption{Achievable limits of spectral efficiency with different number of parallel streams.
\label{HSTfig4}}
\end{figure}

According to \cite{HST22}, totally 30 paths are generated for the 20000 timeslot periods, as shown in Fig.~\ref{HSTfigA}. It can observed that the number of effective channel paths is changed with the time (i.e., the location of HST). Moreover, due to the existence of LoS, the number of effective channel paths $L$ is no less than 1 for all time. Fig.~\ref{HSTfig3} shows the statistical results of numbers of effective channel paths during a HST traverse. It can be seen that nearly 41$\%$ of locations have 3 effective channel paths, i.e., one LoS path and two NLoS paths. Meanwhile, the locations with 2 effective channel paths, i.e., one LoS and NLoS path, account for the second largest proportion with nearly 33$\%$. The proportion of locations with only LoS path is about 13$\%$ and almost the same as that of locations with 4 effective channel paths, i.e., one LoS path and three NLoS paths. However, the locations with 5 effective channel paths, i.e., one LoS path and four NLoS paths account for the smallest percentage of all, which is less than 1$\%$. Assuming all CSI is known perfectly, the maximum spectral efficiencies, i.e., theoretical spectral efficiency limits, are calculated for all locations with different values of parallel streams at the transceivers. Fig.~\ref{HSTfig4} shows the means and variances of theoretical spectral efficiency limit by varying the number of parallel streams at transceivers. It can be seen that the theoretical limits improve with the number of parallel streams. However, when the number of parallel streams is beyond 3, the improvement is too small to be noticed. The reason is that the path loss of higher-order reflected NLoS paths are very severe and their contribution of received signal energy can be neglected. Considering the complexity and cost of hardware implementation, the number of parallel streams   can be set to be 3 to maximize the system capacity as much as possible, which is also adopted in the following simulations.

To evaluate the performance, it is desired to compare our proposed beam searching algorithm with other algorithms via computer simulations. Unfortunately, there does not exist a beam searching algorithm designed for the mmWave HST communications. Since the brute-force sequential beam searching is the most simply and straightforward, we try to modify it to address the temporal correlation of dominant multipath behaviors in mmWave HST channels. In the modified sequential beam searching algorithm, there is totally $M$ measurements for each TTI and $M$ AoD and AoA pairs are selected for measurement. All AoD and AoA pairs associated with the paths that are used for data transmission in the previous TTI will be selected for measurement again. Then, the remaining available measurements are allocated to the AoD and AoA pairs which are sequentially searched from the whole Tx/Rx angle plane. After the measurement, only $D$ paths with best channel conditions will be chosen from the $M$ measured paths for further data transmission. Under the same simulation environments, our proposed beam searching algorithms are compared with the benchmark modified sequential beam searching algorithm.

\begin{figure}[ht]
\centering
\includegraphics[width=4in]{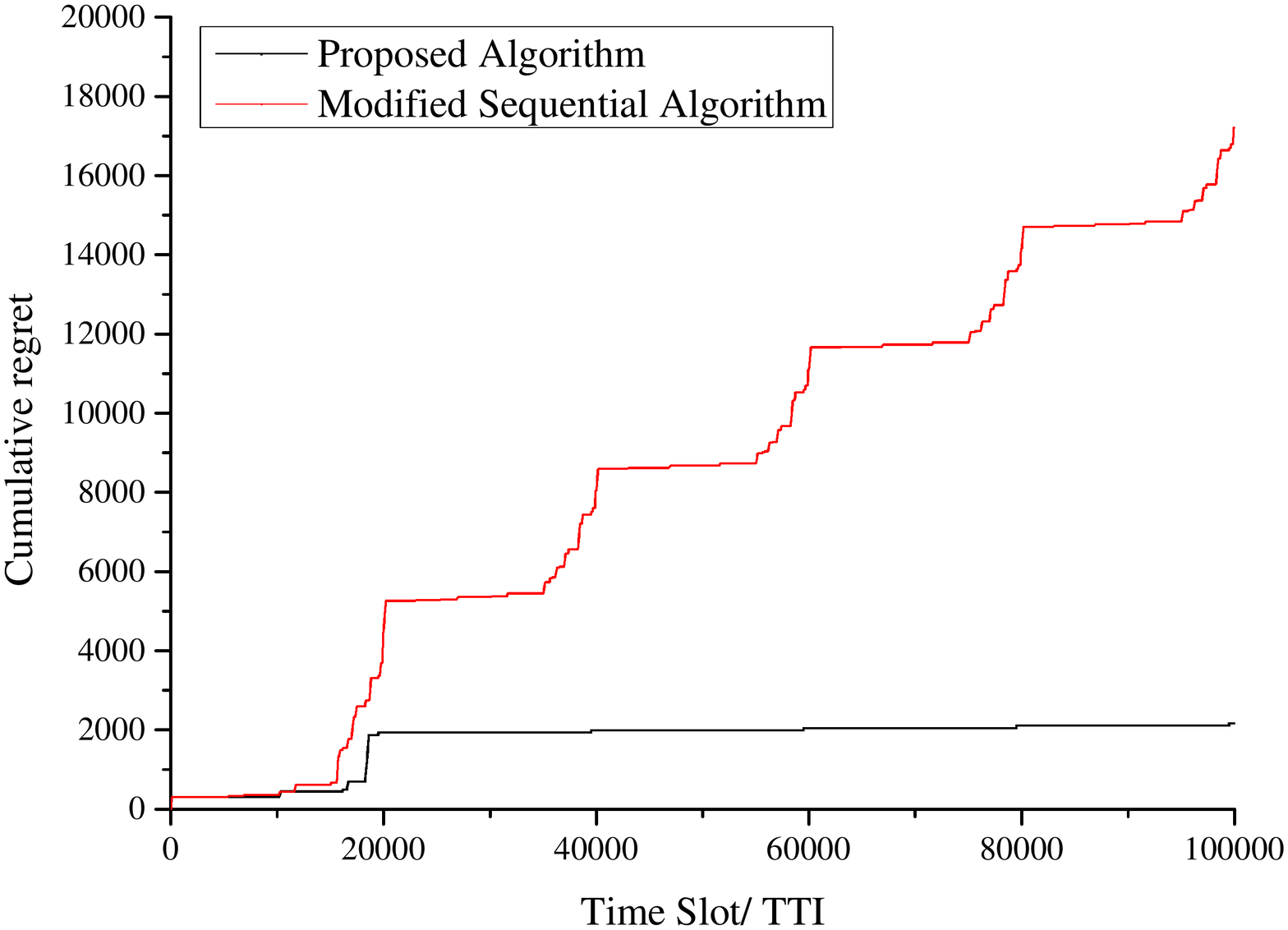}
\caption{Cumulative Loss of Spectral Efficiency (bps/Hz) versus time slot with $M=6$ and $N_{\rm{r}}=N_{\rm{t}}=32$ .
\label{HSTfig5}}
\end{figure}

Fig.~\ref{HSTfig5} compares the cumulative regret of spectral efficiency under the proposed bandit inspired beam searching algorithm and the modified sequential beam searching algorithm. The upper bounds on the proposed bandit inspired beam searching algorithm from Theorem 1 is not plotted here, since the bounds are loose especially. Finding tight upper bounds is a subject of future study. It can be seen from Fig.~\ref{HSTfig5} that the proposed algorithm outperforms the modified sequential algorithm significantly. Moreover, while a HST traverses the coverage of the mRRH for the first time, i.e., from timeslot 1 to 20,000, the regret of the proposed algorithm increases quickly. The reason is that the propagation environment information is unknown in advance. Limited by the number of measurements of each timeslot, the proposed algorithm is searching the all available paths gradually, which results in the loss of achievable spectral efficiency. Meanwhile, the information of all available paths at all locations are obtained and collected with the aid of reward mechanism in the bandit model. Then, when the HST traverses for the second time, i.e., from timeslot 20,001 to 40,000, the regret of the proposed algorithm increases quite small, which indicates nearly all available paths at all locations have been explored and exploited for the data transmissions.

\begin{figure}[ht]
\centering
\includegraphics[width=4in]{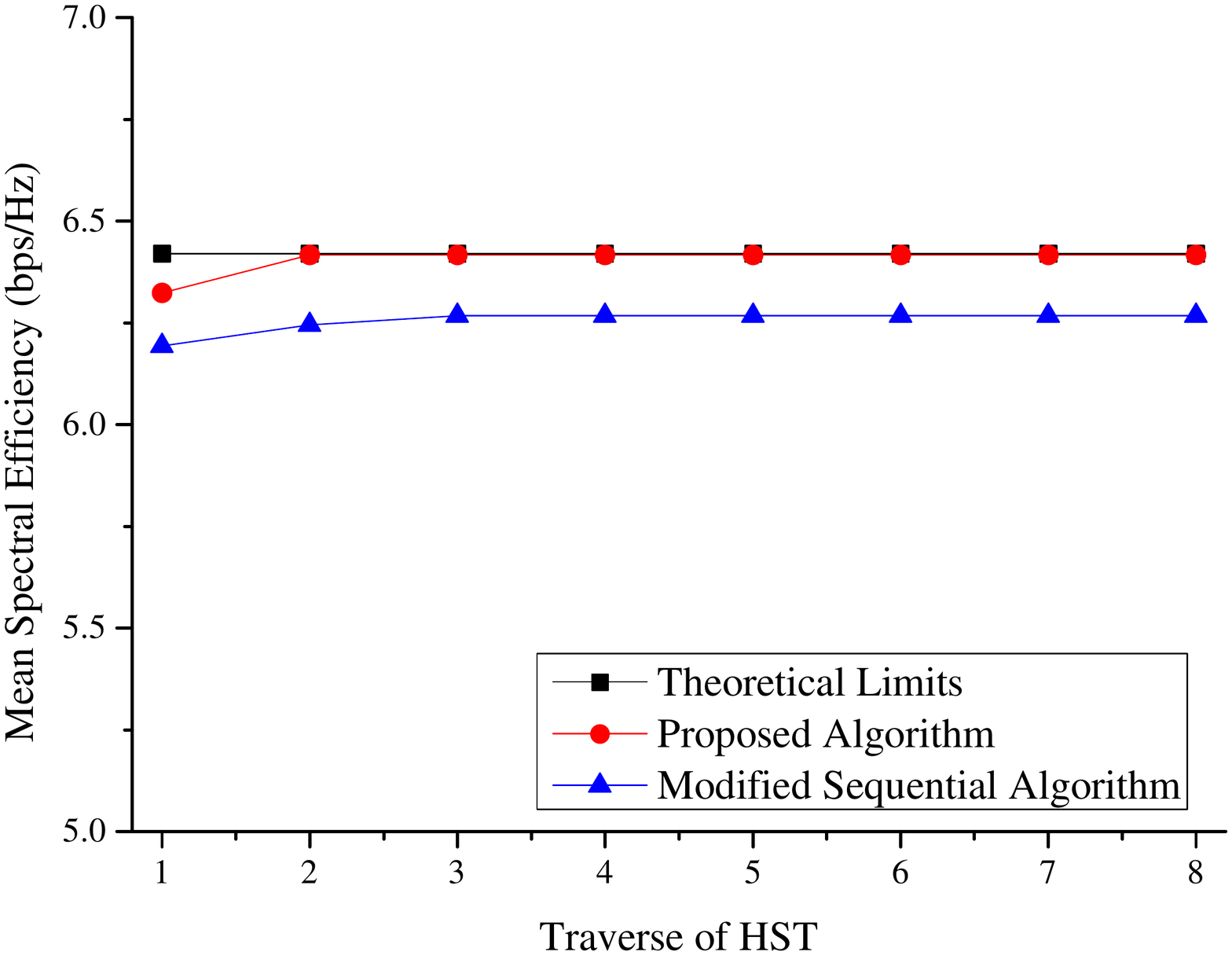}
\caption{Mean Spectral Efficiency (bps/Hz) versus the traverse number of HST with $M=6$ and $N_{\rm{r}}=N_{\rm{t}}=32$.
\label{HSTfig6}}
\end{figure}

In order to further verify the learning effect of the proposed algorithm, Fig.~\ref{HSTfig6} shows the mean spectral efficiency performance versus the traverse number of HST. It can be seen that the mean spectral efficiency performance approaches the theoretical spectral efficiency limits very quickly. More specific, when a second HST traverses, the proposed algorithm obtains nearly the same mean spectral efficiency performance as the theoretical spectral efficiency limits, which indicates the proposed algorithm can learn the propagation environments very quickly and is consistent with the results shown in Fig.~\ref{HSTfig5}.

\begin{figure}[ht]
\centering
\includegraphics[width=4in]{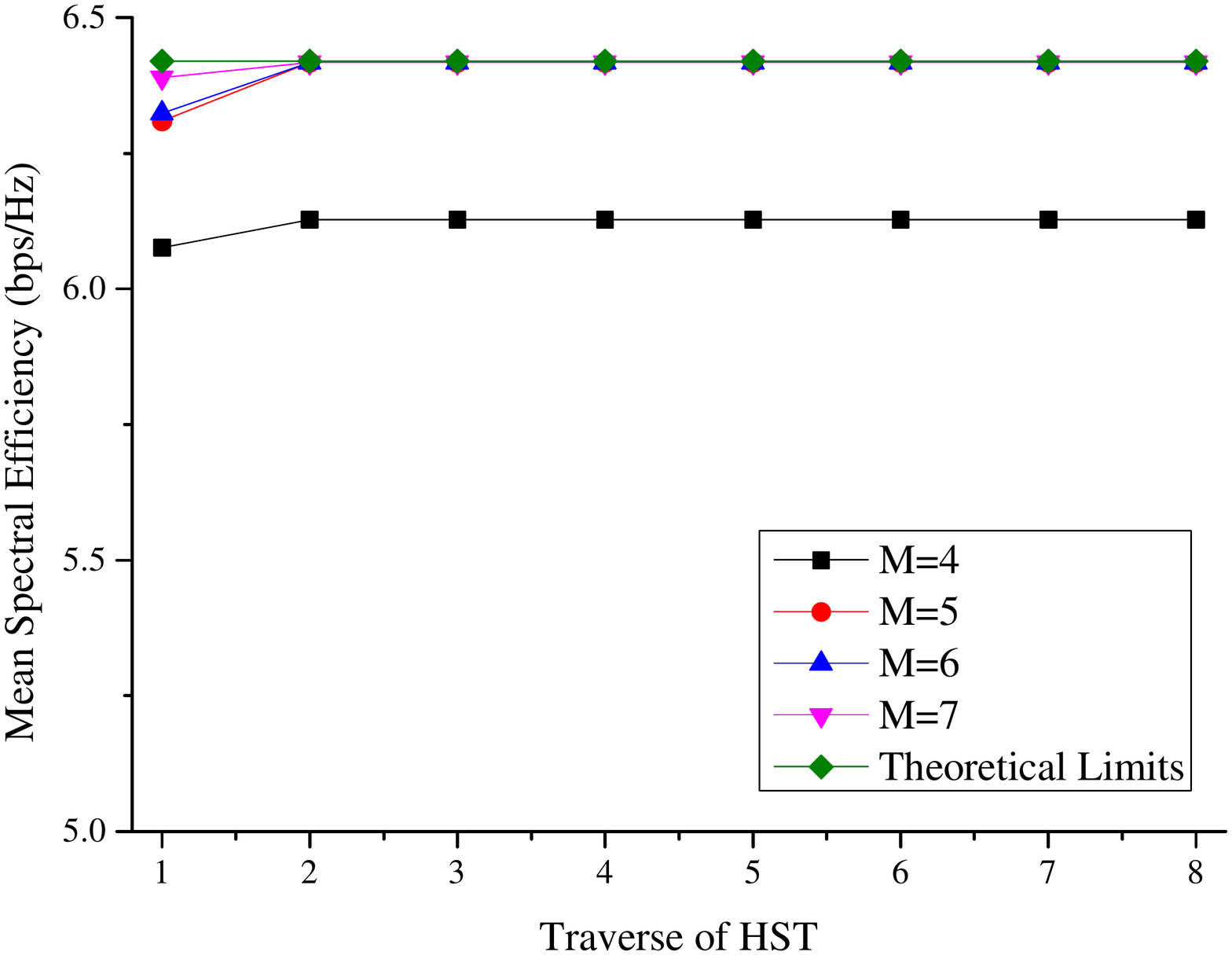}
\caption{Mean spectral efficiency (bps/Hz) versus the traverse number of HST with $N_{\rm{r}}=N_{\rm{t}}=32$ for different number of measurements.
\label{HSTfig7}}
\end{figure}

Fig.~\ref{HSTfig7} shows how the mean spectral efficiency of the proposed algorithm changes with the traverse number of HST for different numbers of measurements. When a first HST traverses, it can be seen that the gap between the proposed algorithm and the theoretical limits decreases with the number of measurements, which indicates increasing the number of measurements will accelerate the exploration of available paths. Moreover, when a second HST traverses, only the proposed algorithm with $M=4$ keeps a gap to the theoretical spectral efficiency limits, which indicates that $M=4$ is too small to explore all available paths quickly. Therefore, an appropriate number of measurements must be selected carefully for practical systems to balance the learning ability of the proposed algorithm and the data transmission capacity.

\begin{figure}[ht]
\centering
\includegraphics[width=4in]{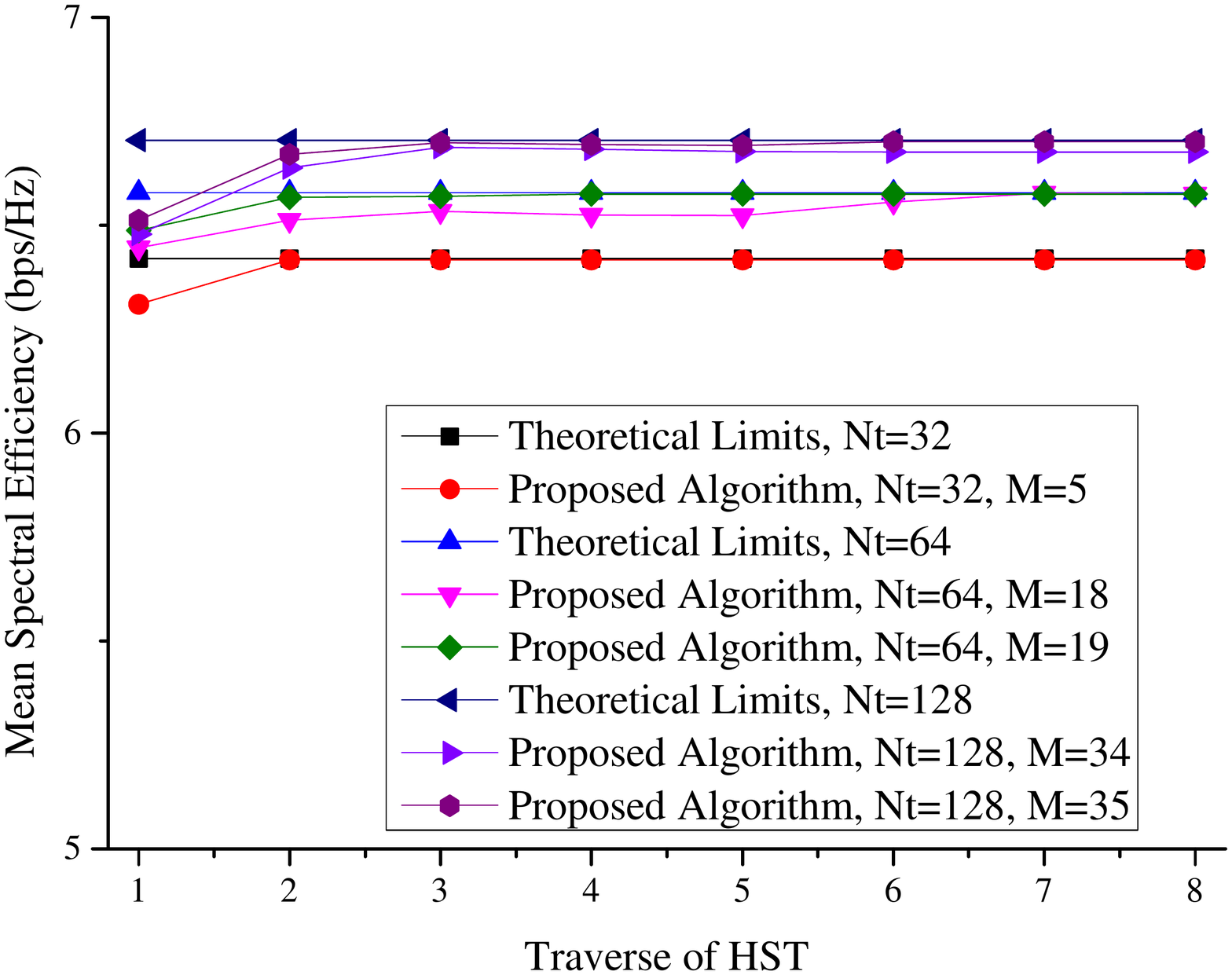}
\caption{ Mean spectral efficiency (bps/Hz) versus the traverse number of HST for different antenna element number of mRRH and different number of measurements.
\label{HSTfig8}}
\end{figure}

Fig.~\ref{HSTfig8} shows the performance of the proposed algorithm with different antenna element number of mRRH and different number of measurements. It can be seen that the theoretical limits of mmWave HST channel increase with the antenna elements of mRRH. The reason is that increasing the elements number of antenna array will narrow the beamwidth and improve the beam gain, which can enhance the received signal strength and reduce the interference among different paths. However, with the growth of the elements number of antenna array, the size of VCM $\bf{H}_{\rm{V}}$ increases proportionally, which results in the enhancement of difficulties for beam searching in mmWave HST channels. Therefore, the number of measurements increases accordingly. However, compared with the number of entries in $\bf{H}_{\rm{V}}$, i.e. $N_{\rm{t}} N_{\rm{r}}$, a very small number of measurements can be selected for the proposed algorithm to approach the performance of the theoretical limits within only a few traverses of HSTs.

\section{Conclusions}
\label{section7}

Since the channel conditions vary quickly in mmWave HST communication systems, channel estimations should be performed frequently. However, conventional channel estimation schemes are not applicable in the HST scenarios. Motivated by the successful applications of the MAB in solving the sequential and selection problems, this paper has formulated the beam searching of mmWave HST communication as a MAB problem and proposed a bandit inspired beam searching scheme to accelerate the channel estimation process. In the proposed scheme, the wireless propagation historic information is exploited to reduce the number of path measurements as much as possible. In addition, the regret of the proposed scheme, which grows logarithmical in time and linearly with the number of arms, was proved to be upper bounded. Finally, simulation results have validated the effectiveness of the proposed scheme, which can approach the theoretical limit with only a few traverses of HSTs.

%In reinforcement learning, the sequential and selection problems can be solved by the multi-armed bandits. This paper has studied the beam searching problem for HST communication systems operating in mmWave bands and proposed a bandit inspired beam searching scheme. The historical channel information can be exploited to assist the channel estimation. We have first formulated the beam searching problem of mmWave HST systems as a MAB problem with each beam as an arm. Then, a bandit inspired scheme has been proposed to reduce the path measurements and accelerate the channel estimation processes. In the proposed scheme, each time slot is divided to exploit the paths which have presented before and explore the paths which have not appeared. In addition, the regret of the proposed scheme, which grows logarithmical in time and linearly with the number of arms, is proved to be upper bounded. Finally, simulation results have validated the effectiveness of the proposed scheme, which can approach the theoretical limit with only a few traverses.

%\section*{Acknowledgements}
%This work is supported by the National Nature Science Foundation of China (No. 61720106003, No. 61571115, No. 61602235), the European Union's Horizon 2020 research and innovation programme under the Marie Sk\l{}odowska-Curie grant agreement (No.709291), and the Natural Science Foundation of Jiangsu Province of China (No. BK20161007).

\bibliographystyle{IEEEtran}
\bibliography{HSTBibliography}

% that's all folks
\end{document}